\documentstyle[sprocl]{article}

\def\IR{{\bf R}}
\def\ZZ{{\bf Z}}
\def\IC{{\bf C}}
\def\IP{{\bf P}}
\def\e{{\bf e}}
\def\pd{\partial}
\def\O{{\cal O}}
\def\({\left(}
\def\){\right)}
\def\tz{\theta_z}
\def\tx#1{\theta_{x_{#1}}}
\def\ta#1{\theta_{a_{#1}}}
\def\da#1{{\pd \; \over \pd a_{#1}}}
\def\D{\Delta}
\def\Ds{\Delta^*}
\def\SD{{\Sigma(\D(w))}}
\def\SDs{{\Sigma(\Ds(w))}}
\def\ns#1{\nu^*_{{#1}}}
\def\bns#1{\bar\nu^*_{{#1}}}
\def\A{{\cal A}}
\def\cp#1{\langle #1 \rangle}
\def\dualp#1{\langle #1 \rangle}
\def\sqbox{{\cal D}}
\def\drho#1{\pd_{\rho_{#1}}}

\def\be{\begin{equation}}
\def\ee{\end{equation}}
\def\bea{\begin{eqnarray}}
\def\eea{\end{eqnarray}}

\begin{document}
\title{GKZ HYPERGEOMETRIC SYSTEMS AND APPLICATIONS TO MIRROR SYMMETRY }
\author{ S. HOSONO }
\address{ Department of Mathematics, Toyama University, \\
         Toyama 930, Japan}
\author{ B.H. LIAN}
\address{ Department of Mathematics, Brandeis University \\
         Waltham, MA 02154}
\maketitle\abstracts{
We analyze GKZ(Gel'fand, Kapranov and Zelevinski) hypergeometric
systems and apply them to study the 
quantum cohomology rings of Calabi-Yau
manifolds. We will relate properties of the local solutions
near the large radius limit to the intersection rings of a toric
variety and of a Calabi-Yau hypersurface. 
 }

\section{Introduction}

Mirror symmetry of Calabi-Yau manifolds is one of the most beautiful 
aspects of string theory
\cite{candelasETAL}. It has been applied with great success to 
do non-perturbative
calculation of quantum cohomology rings\cite{candelasETAL}$^{-}$\cite{HLY}. 
More recently, new
ideas have been developed 
to apply mirror
symmetry to study  the moduli space of the
type II string vacua compactified on a Calabi-Yau manifold.  
Some of the recent work on verifying the so-called
heterotic-type II string duality relies heavily on these new 
ideas\cite{KachruVafa} \cite{KLM} \cite{KKLMV}.
 
One of the
key ingredients for studying families of
Calabi-Yau manifolds is the so-called 
Picard-Fuchs equations. They are
differential equations which govern the period integrals of a Calabi-Yau
manifold. In this report, we will review several aspects of the
Picard-Fuchs equations which arise in mirror symmetry. 
We define the flat coordinates and use them to give a natural 
description of the quantum cohomology ring. 
We relate the flat coordinates to the general solutions 
of the Picard-Fuchs equations at the so-called
point of maximally unipotent
monodromy\cite{Morrison}.  
The general solutions turn out to be in a subspace of the solutions
to
a Gel'fand-Kapranov-Zelevinski (GKZ) 
hypergeometric system\cite{GKZ}.
A GKZ system is therefore reducible in our case. 

\section{ Mirror symmetry and quantum cohomology ring }

To see the essence of the mirror symmetry, we should
go back to 
a toy model ($c=3$) of a string theory 
with target space given by a complex 1-torus 
$T$. We can write it as 
$T=\IC/\Lambda$ where
$\Lambda=\ZZ \e_1 \oplus \ZZ \e_2$ is
a rank 2 lattice in $\IC$. We introduce the metric and the
antisymmetric tensor on the torus by $G_{ij}:=\e_i\cdot\e_j$ and
$B_{ij}=B \epsilon_{ij}$. Then it is natural
to introduce the complexified 
K\"ahler modulus $\lambda=2(B+i\sqrt{G})$ 
and the complex structure modulus
$\rho={G_{12} \over G_{22}} + i {\sqrt{G} \over G_{22}}$. The partition
function $Z_{T}(\lambda,\rho)$ is determined exactly as a function of
moduli and is known to have the following 
symmetries\cite{Dikgraaf}\cite{LercheWarner}: 
$1)Z_{T}(\lambda,\rho)= Z_{T}(\lambda,\rho+1)
                        =Z_{T}(\lambda,-1/\rho)$, 
$2) Z_{T}(\lambda,\rho)=Z_{T}(-\bar\lambda,-\bar\rho)$,  
$3) Z_{T}(\lambda,\rho)=Z_{T}(\rho, \lambda)$.   
The first two symmetry is the result of modular invariance
and orientation invariance. The last symmetry, however, 
comes from the invariance under the
exchange of momenta and winding numbers.
It is the simplest example of mirror symmetry:
an invariance under the exchange of the complexified
K\"ahler moduli and the complex structure moduli.  

If we combine the symmetries $1)$ and $3)$, we may derive the
relation $Z_{T}(\lambda,\rho)=Z_{T}(-1/\lambda,\rho)$, which is
the complex analogue of
the famous symmetry, $R \leftrightarrow 1/R$ duality found by Kikkawa and
Yamazaki\cite{KikkawaY}\cite{Sakai}. 

When the complex 1-torus is replaced by
a Calabi-Yau 3-fold $M$, the symmetry $3)$
becomes a little more involved. First we need a new
manifold $W$ which is "mirror" to $M$. Then we have 
\begin{equation}
Z_M(f(\rho),g(\lambda))=Z_W(\lambda,\rho) \;\; ,
\label{eq:Zmw} 
\end{equation}
where the parameters $\lambda, \rho$ represent local 
coordinates of the appropriate moduli spaces.
The functions $f,g$ are mirror maps which we will later describe 
in terms of the flat coordinates of the Gauss-Manin system. 

There are two local operator algebras, which are called type A,B
respectively,
associated to a string
model compactified along $M$.
The type A algebra is associated to the K\"ahler 
deformation $\lambda$ and it receives  non-perturbative 
quantum corrections from the $\sigma$-model instantons. The type B algebra 
depends on the complex structure but receives no quantum 
correction\cite{DistlerGreene}. 
Then the symmetry
expressed by (\ref{eq:Zmw}) implies that
there is an isomorphism between the
type A algebra of $M$ and the type B algebra of $W$ and
vice versa. This turns out to be a very
powerful tool for computing the quantum correction appearing
in the type A algebra, say, of $M$. This is
the quantum cohomology ring of $M$. 

The type A and B algebras may be described in more geometrical terms
as follows.
In classical geometry, the (complexified) K\"ahler moduli of a  
Calabi-Yau manifold $M$ can be described in terms of
the cohomology group
$H^{1,1}(M,\IC)$, regarded as a subspace of the commutative
algebra $\oplus_{i=0}^3
H^{i,i}(M,\IC)$. On the other hand,
the complex structure deformation of the mirror $W$ may be described
by the
variation of the Hodge structures, for which the space $H^3(W,\IC)=
\oplus_{i=0}^3
H^{3-i,i}(W,\IC)$ and its Hodge filtrations play an essential role.
When $M$ is a  Calabi-Yau hypersurface in a weighted projective space 
$\IP^4(w)$,  the complex structure deformations(, at least 
the algebraic ones,) for the
mirror $W$ can be represented by polynomial
deformations of the defining equation $P(z)$ of $W$. 
In fact we have an isomorphism to the Jacobian ring; $\oplus_{i=0}^3
H^{3-i,i}(W,\IC) \cong \IC[z_1,\cdots,z_5]/( \pd P(z) / \pd z_i )$.
Now mirror symmetry predicts that the quantum cohomology ring of $M$ 
is given by
\begin{equation}
\oplus_{i=0}^3 H^{i,i}_q(M,\IC) \cong \oplus_{i=0}^3 H^{3-i,i}(W_\psi,\IC)\;\;,
\label{eq:qch}
\end{equation}
where $\psi=(\psi_1,\cdots,\psi_{h^{2,1}(W)})$ are
parameters in the polynomial
deformation of the mirror $W$, and the right hand side
is given the structure of the Jacobian
ring $J_\psi$ above. 

In their original work, Candelas et al\cite{candelasETAL} 
determined the quantum cohomology
ring starting from the Jacobian ring $J_\psi$ for the mirror of a 
quintic hypersurface in $\IP^4$. In general we may define the
quantum cohomology ring through a specific basis (flat coordinate)
of the Jacobian ring, $\{ 1, \O_a, \O^b, \O^{(3)} \} \;
(a,b=1,\cdots, h^{2,1}(W))$, where $\O_a, \O^b$ and $\O^{(3)}$ represent 
the elements with charge one, two and three, respectively, in the 
Jacobian ring. The flat coordinate is characterized 
by the properties that
$\O_a \O^b \equiv \delta_a^b \O^{(3)}\;,\; \O_a \O^{(3)}\equiv 0 $ in
the Jacobian ring. Then the relations $\O_a\O_b \equiv \sum_c 
K_{t_at_bt_c}(t(\psi)) \O^c $ determines the coupling
as a function in flat coordinates. We may compute 
the quantum cohomology ring at the so-called large radius
limit where we have non-trivial $q$-expansion for the coupling:
\begin{equation}
K_{t_a t_b t_c}(t)\sim \int_M h_a\wedge h_b \wedge h_c +
\sum N_{i_1 i_2 \cdots i_n} q_1^{i_1}q_2^{i_2} \cdots q_n^{i_n} \;. 
\label{eq:qseries}
\end{equation}
The first term in the expansion are the classical intersection
numbers for the elements $h_a \in H^{1,1}(M,\ZZ) \; (a=1,\cdots,h^{1,1}(M))$
and $q_i:={\rm e}^{2\pi i t_i} \, (i=1,\cdots,n=h^{2,1}(W))$. 
It has been verified in numerous examples that the numbers 
$N_{i_1i_2\cdots i_n}$ are integers, possibly negative, which
"count" the instantons appearing in 
the quantum correction\cite{candelasETAL}$^-$\cite{HLY}.

\section{Gauss-Manin system and flat coordinates}

In this section we will characterize the flat coordinates 
through the analysis of the Gauss-Manin system. 

Let us consider the quintic hypersurface $M$ in $\IP^4$ and its mirror 
$W$ obtained by orbifoldizing $M$ by $G=(\ZZ_5)^3$. We take the
defining equation for $W_\psi$ as
$P_\psi={1\over5}z_1^5+\cdots+{1\over5}z_5^5 -\psi \, \phi$ with
$\phi=z_1z_2z_3z_4z_5$. The deformed Jacobian ring $J_\psi$ is given by 
\begin{equation}
J_\psi:=
\IC[z_1,\cdots,z_5]^G/(\pd P_\psi / \pd z_i)\cap \IC[z_1,\cdots,z_5]^G \,.
\label{eq:qjpsi}
\end{equation}
We fix a basis of $J_\psi$ as
$\{\varphi^{(0)},\varphi^{(1)},\varphi^{(2)},
\varphi^{(3)}\}:=\{1,\phi,\phi^2,\phi^3 \}$ indicating the charges by 
the superscripts (, $\varphi^{(i)}$ refers to the element with
charge $i$ or homogeneous degree $5i$). Then the Gauss-Manin system
is a set of first order differential equations satisfied by the
period integrals. They are given by $w:=(w^{(i)}_{\;\;j})$ with
\begin{equation}
w^{(i)}_{\;\;j}=i!\int_{\gamma_j} Res_{P_\psi=0}(
{ \varphi^{(i)} d\mu \over P_\psi^{i+1} } )\;,
\label{eq:qperiods}
\end{equation}
where $\gamma_j$'s are cycles in $H_3(W_\psi,\ZZ)$ and $d\mu:=\sum_k
(-1)^{k+1}z_k dz_1\wedge\cdots \wedge \hat{dz_k} \wedge 
\cdots \wedge dz_5$. We can derive, using  the reduction pole order 
argument, the differential equation satisfied by (\ref{eq:qperiods}) as 
\begin{equation}
{\pd \; \over \pd\psi}w = 
\( \matrix{ 0 & 1 & 0 & 0 \cr
            0 & 0 & 1 & 0 \cr
            0 & 0 & 0 & 1 \cr
     { \psi \over 1-\psi^5} &  {15\psi^2 \over 1-\psi^5} &
   {25\psi^3 \over 1-\psi^5} & {10\psi^4 \over 1-\psi^5} \cr } \) w
\label{eq:qGM}
\end{equation}

We notice that we can
do a degree-preserving 
change of basis on the Jacobian ring by $\varphi^{(i)}
\rightarrow Q_i(\psi)\varphi^{(i)}+\sum_j R_{ij}(z,\psi) \pd P_\psi/\pd
z_j$, where $Q_i(\psi)$ and $R_{ij}(z,\psi)$ are arbitrary.
We may also change the normalization of the defining
equation by an arbitrary function $r(\psi)$: 
$P_\psi \rightarrow r(\psi) P_\psi$. It is easy to see that 
these changes result in a new period $v$ related to the original one by 
$w=M(\psi)v$ with lower triangular matrix $M(\psi)$. By a 
change of local parameter $\psi=\psi(t)$ to the
flat coordinate $t$, the Gauss-Manin system 
(\ref{eq:qGM}) becomes 
\begin{equation}
\begin{array}{rcl}
{\pd \; \over \pd t}v 
& = & \( {\pd \psi \over \pd t}\) 
      (M^{-1} G_\psi M - M^{-1}{\pd \; \over \pd \psi} M)v  \\
& = & \( \matrix{ 0 & 1 & 0          & 0 \cr
                  0 & 0 & K_{ttt}(t) & 0 \cr
                  0 & 0 & 0          & 1 \cr
                  0 & 0 & 0          & 0 \cr } \)v \;,      \\
\end{array}
\label{eq:flat}
\end{equation}
where $G_\psi$ is the connection matrix in the right hand side of 
(\ref{eq:qGM}) and $K_{ttt}(t)$ is a function which will be determined 
by this form of the Gauss-Manin system. 
We can determine the matrix $M(\psi)$ explicitly as
\begin{equation}
M(\psi)=\(
\matrix{ r & 0 & 0 & 0 \cr
         r' & s & 0 & 0 \cr
         r'' & s'+{r' \over r}s & 
         {1 \over s}{C \over 1-\psi^5} &  0  \cr
         r''' & (s'+{r' \over r}s)'+{r'' \over r}s & 
        {1 \over s}\{ \({1\over 1-\psi^5}\)'+{r'\over r}{1\over 1-\psi^5}\} &
         {1\over r}{C \over 1-\psi^5} }  \) \;,
\label{eq:qmat}
\end{equation}
where the functions $r=r(\psi)$ and $s=s(\psi)$ satisfy differential
equations (, with $r'=\pd r /\pd \psi$ e.t.c.),
\begin{equation}
\begin{array}{rcl}
& r''''-{10 \psi^4 \over 1-\psi^5}r'''-{25\psi^3 \over 1-\psi^5}r''
  -{15 \psi^2 \over 1-\psi^5}r'-{\psi \over 1-\psi^5}r =0     \\
& s''-{5\psi^4 \over 1-\psi^5}s'-
 \({5\psi^3 \over 1-\psi^5}+{5\psi^4 \over 1-\psi^5}{r' \over r}
   +{2r'^2 \over r^2}-{3r'' \over r} \) s =0 \;\;.
\end{array}
\label{eq:diffrs}
\end{equation}
The differential equation for $r(\psi)$ coincides with the Picard-Fuchs
equation. The relation (\ref{eq:flat}) determines the coupling and the
flat coordinate $t=t(\psi)$ as follows;
\begin{equation}
K_{ttt}(t)={1\over r^2}{C \over 1-\psi^5}\({\pd \psi \over \pd t}\)^3
\;,\; 
{\pd t \over \pd \psi}={s(\psi) \over r(\psi) }\;.
\label{eq:Kttt}
\end{equation}

The Picard-Fuchs equation satisfied by $r(\psi)$ in  
(\ref{eq:diffrs}) can be arranged to 
\begin{equation}
\{\tz^4-5z(5\tz+4)(5\tz+3)(5\tz+2)(5\tz+1)\} w(z) = 0 \;,
\label{eq:pf}
\end{equation}
where $z={1\over (5\psi)^5}$ and $w(z)=-5\psi r(\psi)$. It is
evident that 
the indices of the Picard-Fuchs equation are 
zero at the large radius limit $z=0$, and
the monodromy becomes maximally
unipotent. All the solutions can be determined by the standard Frobenius
method starting from the series $w_0(z,\rho):=\sum
{\Gamma(5(n+\rho)+1) \over \Gamma(n+\rho+1)^5} z^{n+\rho}$. 
One can verify that the ratio 
$ t(z)={1\over 2 \pi i}{w_1(z) \over w_0(z) }$, 
where 
$w_0(z):=w_0(z,\rho)\vert_{\rho=0}$ and 
$w_1(z):={\pd \; \over \pd \rho}w_0(z,\rho)\vert_{\rho=0}$,
coincides with the flat coordinate.

In fact the functions 
$r(\psi)=-{1\over 5\psi}w_0(z)$ and 
$s(\psi)=-{1\over 5\psi } w_0 {\pd t \over \pd \psi}$ solve the 
equations (\ref{eq:diffrs}).  Because of the behavior 
$t\sim {1\over 2\pi i}log(z)$ near the large radius limit, 
we obtain the desired $q$-expansion 
(\ref{eq:qseries}) with $q=e^{2\pi i t}$ and $C={5^2 \over (2\pi i)^3}$.

We can read off the prepotential $F(t)$ for the quantum coupling $K_{ttt}$ 
{}from the form of the Gauss-Manin system in the flat coordinate
(\ref{eq:flat}). To see this note that 
the first order system (\ref{eq:flat}) is
equivalent to $\pd_t^2{1\over K_{ttt}(t)}\pd_t^2 v^{(0)}=0$, where
$v^{(0)}$ is the first row of the periods $v=( v^{(i)}_{\;\;j})$. 
Then it is
easy to see that the following $v^{(0)}_{\;\;j}$'s constitute the solutions; 
$v^{(0)}_{\;\;0}=1, 
 v^{(0)}_{\;\;1}=t, 
\pd_t^2 v^{(0)}_{\;\;2}=K_{ttt}, 
\pd_t^2 v^{(0)}_{\;\;3}=-t\,K_{ttt}$. These  
relations are sufficient to defermine the prepotential
\begin{equation}
F(t):={1\over2}\( v^{(0)}_0 v^{(0)}_3+v^{(0)}_1 v^{(0)}_2 \) \;,
\label{eq:qprepot}
\end{equation}
such that $K_{ttt}(t)=\pd^3_t F(t)$. Since we have 
a relation $v^{(0)}_{\;\;j}={1\over r(\psi)} w^{(0)}_{\;\;j}$
{}from the matrix $M(\psi)$ in (\ref{eq:qmat}), we can write the 
prepotential in terms of the solutions of the Picard-Fuchs equation.

\section{ GKZ hypergeometric system and the flat coordinate }

In this section, we will consider a Calabi-Yau hypersurface $X_d(w)$ in 
a weighted projective space $\IP^4(w)$ (, where $d$ represents the 
homogeneous degree of the surface). We will consider two a priori 
different objects: the intersection ring for 
$X_d(w)$ and 
the GKZ hypergeometric system for the periods of $X^*_d(w)$. We will
find a close relationship between the two. 

\subsection{Intersection ring}

According to Batyrev\cite{BatyrevI}, we can construct the mirror pairing 
$(M,W)=(X_d(w),$ $X_d^*(w))$ of Calabi-Yau manifolds starting from a 
pairing of the reflexive polyhedra 
$(\D(w),\Ds(w))$ in $\IR^4$. Here the polyhedron $\D(w)$ is defined
through the Newton polyhedron of the defining equation (, the potential,)
of $X_d(w)$ and may be written as
\begin{equation}
\D(w)={\rm Conv.}\( \{ \; x \in \ZZ^5 \,\vert\, 
\sum w_i x_i =0 \,,\, (x_i \geq -1)\; \} \).
\label{eq:delta}
\end{equation}
We see that all vertices of this polyhedron are integral by definition and
the origin is the only interior integral point in $\D(w)$
(, in fact this is the defining property of the reflexive polyhedron).
The (polar) dual of $\D(w)$ is defined by $\Ds(w):=\{\, y\in
\Lambda(w)^*_\IR \, \vert \, \langle y, x \rangle \geq -1 \;\}$ and
turns out to be reflexive if $\D(w)$ is reflexive.
Here $\Lambda(w)_\IR$ is the scalar extension of the lattice 
$\Lambda(w)=\{\; x\in \ZZ^5 \,\vert\, \sum w_i x_i =0 \;\}$.
The cones over the faces of the respective polyhedra define the complete
fans $\SD$ and $\SDs$. These fans define compact toric
varieties $\IP_\SD$ and $\IP_\SDs$. Then the Calabi-Yau hypersurfaces 
$X_d(w)$ and $X_d^*(w)$ are crepant resolutions of zero loci of certain 
Laurent polynomials in the ambient spaces $\IP_\SDs$ and $\IP_\SD$,
respectively. 

If  all singularities of the ambient space $\IP_\SDs$ are Gorenstein,
the subdivision of $\SDs$ using all integral points on the faces makes
$\SDs$ regular and results in the smooth ambient space $\IP_\SDs$. This
is the case for the models of type I and II in the classification given 
in refs\cite{HKTY}\cite{HLY}. 
For simplicity, we will restrict our attention to this case. 

The classical cohomology $\oplus H^{i,i}(X_d(w),\ZZ)$ 
may be understood as the restrictions of the cohomology of the ambient
space $\IP_\SDs$. The cohomology ring of the smooth toric variety is 
generated by the toric divisors associated to each one dimensional cones.
We denote the integral points in $\Ds$ as $\ns{1},\cdots,\ns{p}$ and the
corresponding divisors as $D_1,\cdots,D_p$. Then the cohomology ring
called an intersection ring, of the toric variety can be described
\cite{Oda} by 
\begin{equation}
A^*(\IP_\SDs,\ZZ) 
= \ZZ[D_1,\cdots,D_p]/(SR_\SDs+I) \;\;,
\label{eq:Aint}
\end{equation}
where $SR_\SDs$ is the Stanley-Reisner ideal for the fan $\SDs$ and $I$
represents the divisors of the rational functions. These two ideals are
generated, respectively, by 
\begin{list}{}{}
\item[i)] $D_{i_1}\cdots D_{i_k}$ for $\ns{i_1}\cdots\ns{i_k}$ not in a
cone of $\SDs$,
\item[ii)] $\sum_{i=1}^p \langle u,\ns{i} \rangle \, D_i$ for $u\in \ZZ^4$.
\end{list}
\vskip-1cm \begin{equation} \label{eq:ideal} \end{equation}
\vskip0.5cm
The degree four element in $A^*(\IP_\SDs,\ZZ)$ should give the 'volume'
form of $\IP_\SDs$. Since we know that
the total Chern class is $c(T_{\IP_\SDs})=\prod_{i=1}^p (1+D_i)$,
and
$\chi(\IP_\SDs)=$
\# of 4-dimensional cones in $\SDs$\cite{Oda}, 
we can normalize the volume form 
using the relation $\int_{\IP_\SDs}c(T_{\IP_\SDs})=\chi$. With this
normalization, we may define the intersection couplings 
\begin{equation}
\cp{D_{i_1}D_{i_2}D_{i_3}D_{i_4}}:=
\int_{\IP_\SDs} J_{D_{i_1}} J_{D_{i_2}} 
                J_{D_{i_3}} J_{D_{i_4}.}
\label{eq:cpamb}
\end{equation}

Since the divisor of the hypersurface has an expression
$[X_d(w)]=D_1+\cdots+D_p$, the intersection ring of the Calabi-Yau 
hypersurface $X_d(w)$ may be described by
$A^*(X_d(w),\ZZ)=A^*(\IP_\SDs,\ZZ)/Ann(D_1+\cdots+D_p)$, where $Ann(x)$
consists of those elements which vanish after multiplication by $x$ in
the ring $A^*(\IP_\SDs,\ZZ)$. Then the intersection coupling of $X_d(w)$
may be written as $\cp{D_{i_1} D_{i_2} D_{i_3} [X_d(w)]}$. 

\subsection{GKZ hypergeometric system}

As remarked by Batyrev in \cite{BatyrevII}, 
a period integral of $X^*_d(w)$
satisfies the GKZ hypergeometric system. 
This system is defined by the integral points 
$\A=\{ \, \bns{i}=(1,\ns{i}) \,\}_{i=0,\cdots,p}$ in $\Ds$ placed on a
hyperplane in $\IR^5$, and an exponent $\beta \in \IR^5$. 
The set $\A$ is not linearly independent but 
has affine relations expressed by a lattice
$
L=\{\, (l_0,\cdots,l_p)\in \ZZ^{p+1}\, \vert \,\sum_{i} l_i \bns{i} =0 \,\} \;.
$
Then the GKZ hypergeometric system for the period integral $\Pi(a)$ is
given by 
\begin{equation}
\sqbox_l \Pi(a)=0 \;(l\in L)\;\;,\;\; {\cal Z}_u \Pi(a)=0 \;(u\in \ZZ^5)\;,
\label{eq:gkz}
\end{equation}
where
\begin{equation}
\sqbox_l =\prod_{l_i>0}\( \da{i} \)^{l_i} - \prod_{l_i<0}\( \da{i} \)^{l_i} 
\;,\; 
{\cal Z}_u=\sum_i \dualp{u,\bns{i}}\ta{i} - \beta_u \;,
\label{eq:boxop}
\end{equation}
where $\beta_u=\dualp{u,\beta}$ with $\beta=(-1,0,0,0,0)$. This system
has a formal solution
\begin{equation}
\Pi(a,\gamma)=\sum_{l\in L} {1\over \prod_{0\leq i\leq p}
\Gamma(l_i+\gamma_i+1)} a^{l+\gamma} \;\;,
\label{eq:formalsol}
\end{equation}
for $\gamma \in \IR^{p+1}$ satisfying $\sum_i \gamma_i \bns{i} =\beta$. 
It is shown in \cite{GKZ} that convergent powerseries solutions can be 
constructed from this formal solution for each regular triangulation 
of the polyhedron $P={\rm Conv.}\( \{ 0,\bns{0},\cdots,\bns{p} \}\)$.
The regular triangulation determines local variables
$x_k=a^{l^{(k)}}$ through a compatible basis 
$\{ l^{(1)},\cdots,l^{(p-4)} \}$ of the lattice $L$.

The $q$-series expansion of the quantum coupling
(\ref{eq:qseries}) is valid at the large radius limit where the
monodromy becomes maximally unipotent. There is in fact a regular
triangulation which realizes this property. It is a triangulation 
for which we have 
only one power series solution and all other
solutions have logarithmic singularities. This regular
triangulation has the property that  
all 4-simplices contain the point $\bns{0}$(,the origin in $\Ds$,) 
and have volume one(,where
the volume should be normalized so that the unit 'cubic' in
$n$-dimensions has volume $n!$). In \cite{HLY}, we called the regular
triangulation with this property the maximal triangulation. 
Note that the maximal triangulation is a triangulation we can associate
with the complete fan $\SDs$ and thus with the desingularization of the
ambient space $\IP_{\SDs}$. 

The relation between the maximal triangulation $T_0$ 
of the polyhedron $P$ and the ambient space
$\IP_\SDs$ has an important  implication on the
quantum cohomology. To see this, let us write the power series solution 
$w_0(x,\rho)=a_0\Pi(a,\gamma)$ in terms of the integral basis
$\{l^{(1)}, \cdots, l^{(p-n)} \}$ of $L$ compatible with $T_0$, where we
define the indices $\rho$ by $\gamma=\sum_k \rho_k
l^{(k)}+(-1,0,\cdots,0)$ and $x_k=(-1)^{l^{(k)}_0} a^{l^{(k)}}$.
Explicitly, this series has the form
\begin{equation}
w_0(x,\rho)=\sum_{m_1,\cdots,m_k \geq 0}
{\Gamma(-\sum((m_k+\rho_k)l_0^{(k)}+1) \over
 \prod_{1\leq i \leq p} 
 \Gamma(\sum(m_k+\rho_k)l_i^{(k)}+1) } x^{m+\rho} \;\;.
\label{eq:wnot}
\end{equation}
Because of the difference of the 'gauge' factor $a_0$ between
$w_0(x,\rho)$ and $\Pi(a,\gamma)$, the first order differential operator
${\cal Z}_u$ takes the form $\tilde{\cal Z}_u=\sum_i 
\dualp{u,\bns{i}}\ta{i}$ for $w_0(x,\rho)$. In this form, 
this differential operator coincides with the linear relation ii) in
(\ref{eq:ideal}) under the identification $\ta{i} \leftrightarrow D_i
\, (i=1,\cdots,p)$. Under this identification, the 
generators in i) of (\ref{eq:ideal}) 
also correspond to the leading terms of certain
differential operators $\sqbox_l$ as
follows.
Consider the toric ideal $\dualp{\sqbox_l \,\vert\, l\in L} \subset 
\IC[\da{0},\cdots,\da{p}]$, construct a Gr\"obner basis of
this ideal (, with respect to a term order defined by the maximal 
triangulation), and consider the leading terms of the basis
elements.
After a suitable multiplication by a monomial of the form 
$a^{l_+}$ or $a^{l_-}$ with $l=l_{+}-l_{-}$, the leading term of 
a basis element $\sqbox_l$
becomes the principal part of the operator near
$x_k=0$. The basis elements together determines the
local solutions completely. It can be shown that the principal parts
of the Gr\"obner basis elements coincides with the generators i) in
(\ref{eq:ideal}) for the Stanley-Reisner ideal $SR_\SDs$ under the
identification $\ta{i}$ with $D_i$\cite{HLY}. 

Now we can determine the solutions with logarithmic singularities from
the principal parts of the GKZ system using the Frobenius method. 
We observe that the solutions are given by 
\begin{equation}
\begin{array}{rcl}
&w_0(x,0) \;,\; \drho{i} w_0(x,0) \;,\;
\sum_{k,l}C_{ijkl}\drho{k} \drho{l} w_0(x,0) \;,\; \\
&\sum_{j,k,l}C_{ijkl}\drho{j} \drho{k} \drho{l} w_0(x,0) \;,\;
\sum_{i,j,k,l}C_{ijkl} \drho{i}\drho{j}\drho{k}\drho{l} w_0(x,0) \;,\\
\end{array}
\label{eq:ambsol}
\end{equation}
where $C_{ijkl}:=\cp{\tx{i} \tx{j} \tx{k} \tx{l}}$ is the coupling 
(\ref{eq:cpamb}) under the identification above and 
$\drho{i}:={1\over 2\pi i}{\pd \over \pd \rho_i}$. 

It has been reported\cite{HKTY}\cite{HLY} that our GKZ system 
is reducible and we can extract canonically the solutions to 
a subsystem by considering the restriction of the intersection ring
to the hypersurfaces $X_d(w)$. That is, we use in (\ref{eq:ambsol}) 
the cubic coupling $K^{cl}_{ijk}=\cp{\tx{i}\tx{j}\tx{k}[X_d(w)]}$ 
instead of $C_{ijkl}$ for the ambient space. It has been verified
experimentally that the system with these solutions
coincides with the Picard-Fuchs
equations derived from the reduction argument of Dwork-Griffiths-Katz. 
In terms of the local solutions near the large
radius limit, we can write the prepotential in a concise form\cite{HKTY}  
\begin{equation}
F(t)={1\over2}\({1\over w_0}\)^2 \{ w_0 D^{(3)} w_0 +
\sum_i D^{(1)}_i w_0 D^{(2)}_i w_0 \}_{x_k=x_k(q)} \;,
\label{eq:generalF} 
\end{equation}
where $D^{(1)}_i:=\drho{i}\,,\, D^{(2)}_i:={1\over2}\sum_{j,k}
K^{cl}_{ijk}\drho{j}\drho{k} \,,\,
D^{(3)}:=-{1\over6}\sum_{ijk} K^{cl}_{ijk}\drho{i}\drho{j}\drho{k}$.
The mirror map is then defined by $t_k:={D^{(1)}_k w(x,0) \over w_0(x,0) }$. 
It turns out that the asymptotic form of the prepotential is 
$F(t)={1\over6}\sum_{ijk}K^{cl}_{ijk}t_it_jt_k-
\sum_k {c_2\cdot J_k \over
24}t_k -i {\zeta(3) \over 16\pi^3}\chi(X_d(w)) + {\cal O}(q)$.

\section{Summary} 

We have discussed the flat coordinates of the Gauss-Manin system in the
context of the mirror symmetry. In these coordinates,
we can compute the quantum
cohomology ring $\oplus_i H^{i,i}_q(M,\IC)$ using the
Jacobian ring $J_\psi$ of the mirror. 

In case of the toric realization of the mirror symmetry, 
we can trace the structure of the quantum cohomology ring 
to the logarithmic solutions to the GKZ hypergeometric system near 
the large radius limit. 
The closed formula for the prepotential
(\ref{eq:generalF}) is written completely in terms of 
the data of the
reflexive polyhedron $\Ds$. 
It is an interesting and important problem to
relate the $q$-series expansion of the prepotential
to the axiomatic definition of the quantum cohomology 
ring
\cite{RuanTian}
\cite{KM}.

We thank S.T. Yau for his collaboration.
S.H. is supported in part by Grant-in-Aid for Science
Research on Priority Area 231 "Infinite Analysis". 


\end{document}